\documentclass[10pt, a4paper, fleqn]{article}

\usepackage{amsmath}
\usepackage{natbib}
\usepackage{graphicx}
\usepackage{bm} 
\usepackage{amssymb} 
\usepackage{appendix}
\usepackage[explicit]{titlesec}
\usepackage[a4paper]{geometry} 
\usepackage{setspace}
\usepackage{hyperref}
\usepackage{amsthm} 
\usepackage[all]{hypcap} 

\newtheorem{rmk}{Remark}
\newtheorem{property}{Property}
\newtheorem{result}{Result}

\newcommand{\Bn}{\mathcal{B}} 
\newcommand{\dd}{\mathrm{d}} 

\newcommand{\pib}{\pi_{\mathrm{B}}}
\newcommand{\pia}{\pi_{\mathrm{A}}} 
\newcommand{\dkk}{\mathbf{d}} 
\newcommand{\ck}{(c_0, \ldots, c_n)} 
\newcommand{\ckk}{\bm{c}} 
\newcommand{\rkk}{\bm{r}} 
\newcommand{\rk}{(r_0, \ldots, r_{n+1})} 

\newcommand{\xl}{x_\mathrm{L}} 
\newcommand{\xr}{x_\mathrm{R}} 

\begin{document}

\title{Gains from switching and evolutionary stability in multi-player matrix games}
\author{Jorge Pe\~na$^{1,*}$ \and Laurent Lehmann$^{2}$ \and Georg N\"oldeke$^{3}$}
\date{}

\maketitle

\onehalfspace

\vfill

\begin{itemize}
\item[$^{1}$] Faculty of Business and Economics \\ University of Basel \\ Peter Merian-Weg 6, CH-4002 Basel, Switzerland \\
\textit{Current address:}\\
Research Group for Evolutionary Theory \\ Max Planck Institute for Evolutionary Biology \\ August-Thienemann-Str.~2, 24306 Pl\"on, Germany \\
e-mail: \href{mailto:pena@evolbio.mpg.de}{pena@evolbio.mpg.de}\\
\item[$^{2}$] Department of Ecology and Evolution \\ University of Lausanne \\ Le Biophore,  CH-1015 Lausanne, Switzerland \\
e-mail: \href{mailto:laurent.lehmann@unil.ch}{laurent.lehmann@unil.ch}\\
\item[$^{3}$] Faculty of Business and Economics \\ University of Basel \\ Peter Merian-Weg 6, CH-4002 Basel, Switzerland \\
e-mail: \href{mailto:georg.noeldeke@unibas.ch}{georg.noeldeke@unibas.ch}\\
\item[*] Corresponding author.
\end{itemize}

\newpage

%
%

\begin{abstract}

In this paper we unify, simplify, and extend previous work on the evolutionary dynamics of symmetric $N$-player matrix games with two pure strategies.
In such games, gains from switching strategies depend, in general, on how many other individuals in the group play a given strategy.
As a consequence, the gain function determining the gradient of selection can be a polynomial of degree $N-1$.
In order to deal with the intricacy of the resulting evolutionary dynamics,
we make use of the theory of polynomials in Bernstein form.
This theory implies a tight link between the sign pattern of the gains from switching on the one hand and the number and stability properties of the rest points of the replicator dynamics on the other hand.
While this relationship is a general one, it is most informative if gains from switching have at most two sign changes,
as it is the case for most multi-player matrix games considered in the literature.
We demonstrate that previous results for public goods games are easily recovered and extended using this observation.
Further examples illustrate how focusing on the sign pattern of the gains from switching obviates the need for a more involved analysis.


\end{abstract}


\newpage
\doublespacing


\section{Introduction}

Game theory has been widely applied to evolutionary biology~\citep{MaynardSmith1982,Eshel96,Rousset04,Vincent2005,Broom2013}.
Evolutionary game theory has been instrumental in explaining the evolution of traits as diverse as the sex ratio, dispersal, mate competition, parasite transmission, flowering time, cooperation, policing, dormancy, and anisogamy~(e.g., \citealp{Comins80,MaynardSmith1982,ClarkM86,Frank87,Bulmer94,Frank95,BulmerP02,Foster04,Rousset04,OttoD07}).

Evolutionary models of these traits often assume ``playing the field" type of interactions \citep[p.~23]{MaynardSmith1982},
where the payoff to an individual depends on an average property of the population or the group with which it interacts.
There are also many situations, however, where the payoff to an individual depends critically on the strategy profile in the population (or its group)
and where the actions of different individuals cannot be averaged; that is, mass action does not apply.
Typical examples involve collective action problems in moderately sized groups, where the change in behavior by a single individual can result in a large, discontinuous change in payoffs to others~(e.g.,~\citealp{Boyd1988}). Such collective action problems have been modeled as multi-player (or multi-person) matrix games~\citep{Broom1997,Kurokawa2009,Gokhale2010}. Except for the very special cases in which group size is taken to be equal to two (so that the well-developed theory of two-player matrix games can be applied,
cf.~\citealp{Weibull1995,Hofbauer1998,Cressman2003}) or the payoff structure is linear (as in the standard model of the $N$-person prisoner's dilemma) such games have proven difficult to analyze.

The intrinsic complexity of multi-player matrix games is already evident
for the case of symmetric games with two pure strategies A and B on which we focus in this paper. For these games, the average payoff difference in a large and well-mixed population is given by the
so-called \emph{gain function}~\citep{Bach2006}
\begin{equation*}
g(x) = \sum_{k=0}^{n} \binom{n}{k} x^k (1-x)^{n-k} d_k.
\end{equation*}
Here $n$ is the number of co-players of a focal player (so that $N = n+1$ is the group size),
$x$ is the population fraction of A-strategists,
and $d_k$
is the gain a focal player would obtain if switching from strategy B to strategy A when $k$ other group-members play A. The evolutionary solution of the game
(such as the set of evolutionarily stable strategies, ESSs, or the set of stable rest points of the replicator dynamics)
involves not only finding the roots of the gain function $g(x)$ (a polynomial of degree $n$)
but also determining the behavior of $g(x)$ in the vicinity of such roots.
While this is straightforward for two-player games and multi-player games with a linear payoff structure
(for which $g(x)$ linear in $x$),
it is less evident for more general multi-player games.
Indeed, non-linear payoff structures in groups of size larger than five may lead to
polynomials of degree greater than four that, in general, cannot be solved analytically~\citep{Clark1984}.

In order to deal with such complexity, the vast majority of previous works on multi-player matrix games has
considered particular functional forms for the specification of the payoffs and
has resorted to lengthy algebra or numerical methods to study the models~\citep{Joshi1987,Boyd1988,Dugatkin1990,Weesie1998,
Hauert2006a,Zheng2007,Cuesta2008,Pacheco2009,Archetti2009,Souza2009,Archetti2011,VanSegbroeck2012
}.
This way some non-linear public goods games, including multi-player extensions of well-known two-person matrix games such as the stag hunt~\citep{Skyrms2004}
and the snowdrift game~\citep{Sugden1986}, have been characterized on a case-by-case basis.

In contrast to these efforts,~\cite{Motro1991} and~\cite{Bach2006} have taken a more systematic approach to the study of non-linear public goods games.
Both of these papers consider situations in which each contributor to a public good pays a constant cost,
whereas the benefit from the public good, which is obtained by all players, is a function of the number of contributors.
\cite{Motro1991} proves that in this case the replicator dynamics has at most one interior rest point
if the benefit is concave or convex in the number of contributors.
He also provides necessary and sufficient conditions for the existence of such a rest point and characterizes the stability property of all rest points.
In a similar spirit,~\cite{Bach2006} find sufficient conditions on the shape of the benefits such that there exists a critical cost level with the property that for costs below such level the replicator dynamics has two interior rest points, whereas for higher costs there is no interior rest point.

More recently,~\cite{Gokhale2010} have discussed the relationship between the sign pattern of the gains from switching and the number of interior rest points of the replicator dynamics.
Specifically, these authors observe that the replicator dynamics has a single interior rest point if the sequence $(d_0,d_1,\ldots,d_{n})$,
which we refer to as the \emph{gain sequence},
has exactly one sign change.
\citet{Gokhale2010} also note that the direction of selection (as given by the sign of the gain function $g(x)$)
cannot have more sign changes than the gain sequence. This implies that the number of sign changes of the gain sequence provides an upper bound on the number of interior rest points of the replicator dynamics. The latter observation is also made in \citet{Hauert2006a} and \citet{Cuesta2007}.

In this paper, we show how sign-change conditions like the ones discussed in \citet{Gokhale2010} can be refined by using the fact that the gain function $g(x)$ is a particular kind of polynomial,
known as polynomial in Bernstein form (or Bernstein polynomial) with coefficients given by the gain sequence $(d_0,d_1,\ldots,d_{n})$.
Polynomials in Bernstein form are rich in shape-preserving properties,
long recognized in the fields of approximation theory~\citep{Bernstein1912,Lorentz1986,DeVore1993} and computer aided geometric design~\citep{Yamaguchi1988,Farin2002,Farouki2012}. Our analysis rests on the variation-diminishing property of Bernstein polynomials and a property that we refer to as the preservation of initial and final signs. These properties provide a tight link between the sign pattern of the gain sequence and the sign pattern of the gain function.\footnote{The fact that the gain function $g(x)$ is a Bernstein polynomial has previously been noted by \citet{Cuesta2007}. These authors also suggest that the variation diminishing property of these polynomials may make the analysis of many multi-player games straightforward, but do not pursue this idea.}
In particular, if the gain sequence has at most two sign changes, a full characterization of the possible dynamic regimes is easily obtained.

For most of the collective action problems that have been modeled as multi-player matrix games it is straightforward to determine the sign pattern of the gain sequence.
Moreover, because the gain sequences of these games have at most two sign changes, our characterization results provide all the information necessary to recover the results on the number and stability of rest points obtained in previous studies.
We demonstrate these claims for two classes of public goods games, namely threshold games~\citep[e.g.,][]{Dugatkin1990,Weesie1998,Zheng2007,Souza2009},
and constant cost games~\citep[e.g.,][]{Motro1991,Bach2006,Hauert2006a,Pacheco2009,Archetti2011}, and two additional examples taken from \citet{VanSegbroeck2012} and \citet{Hauert2006a}, thus supporting the claim that the approach developed here unifies, simplifies, and extends much of the previous work on multi-player matrix games.

\section{Model}

Interactions occur in groups of size $N = n+1$, in which a focal individual plays a game against $n$ co-players or opponents.
Each individual can choose between one of two different pure strategies, A and B.
The game is symmetric
so that, from the focal's point of view, any two co-players are exchangeable.

Let $a_k$ denote the payoff to an individual choosing A
when $k$ opponents choose A (and hence $n-k$ co-players choose B);
likewise, let $b_k$ denote the payoff to an individual choosing B
when $k$ opponents choose A.
Also let
\begin{equation*}
 d_k \equiv a_k - b_k
\end{equation*}
denote the gain the focal player makes from choosing A over B, taking the choices of other players
($k$ playing A and $n-k$ playing B) as given. The parameters $d_k$, which describe the gains from switching, are collected in the gain sequence $\dkk = (d_0,d_1, \ldots, d_n)$. We assume $\dkk \not= \bm 0$, thus excluding the trivial and uninteresting case in which payoffs are independent of the actions chosen.

Evolution occurs in an infinitely large and well-mixed population
with groups randomly formed by binomial sampling.
Hence, if the frequency of A-strategists in the whole population is $x$,
the average payoffs obtained by an A-strategist
and a B-strategist are respectively given by
\[
 \pia(x) = \sum_{k=0}^{n} \binom{n}{k} x^k (1-x)^{n-k} a_k
\]
and
\[
 \pib(x) = \sum_{k=0}^{n} \binom{n}{k} x^k (1-x)^{n-k} b_k .
\]

We assume that the rules of transmission of the strategies
(whether genetically encoded or individually or socially learned) are such that the frequency $x$ of A-strategists in the population can be described by the replicator dynamics~\citep{Taylor1978,Hofbauer1998}
\begin{equation}
\label{eq:rd}
 \frac{\dd x}{\dd t} = x(1-x)g(x),
\end{equation}
where $g(x) = \pia(x) - \pib(x)$ is the gain function~\citep{Bach2006} given by
\begin{equation}
\label{eq:gx}
 g(x) = \Bn_n(x; \dkk) \equiv \sum_{k=0}^{n} \binom{n}{k} x^k (1-x)^{n-k} d_k.
\end{equation}
As we have already mentioned in the Introduction, the gain function is a \emph{polynomial in Bernstein form} (also known as a \emph{Bernstein polynomial}, cf.~\citet{Farouki2012}).
This is made explicit by the notation we introduce in~\eqref{eq:gx}, where the \emph{Bernstein operator} $\Bn_n$ maps the vector of \emph{Bernstein coefficients} $\dkk \in \mathbb{R}^{n+1}$ into the polynomial $\sum_{k=0}^{n} \binom{n}{k} x^k (1-x)^{n-k} d_k$ in the variable $x \in [0,1]$.

The replicator dynamics~\eqref{eq:rd} has two trivial rest points at $x=0$
(where the whole population consists of B-strategists) and $x=1$
(where the whole population consists of A-strategists).
Interior rest points $0 < x^* < 1$ are given by the solutions of the equation $g(x^*) = 0$.
Because $g(x)$ is a polynomial of degree at most $n$ (and we have assumed $\dkk \not = \bm 0$)
the replicator dynamics can have at most $n$ interior rest points, corresponding to $n$ simple roots of $g(x)$ in the open interval $(0,1)$.
In the two-strategy case we analyze here, rest points of the replicator dynamics can be either (locally asymptotic) stable or unstable. Stability of a rest point $x^*$ requires that $(x-x^*)(g(x) - g(x^*)) < 0$ holds for all $x \not= x^*$ in a neighborhood of $x^*$.
Since the stable rest points of the replicator dynamics correspond to ESSs for the multi-player game~\citep{Bach2006}, our following results about stable rest points of the replicator dynamics carry over to ESSs without any changes.

\begin{rmk}
 The gain function $g(x)$ given in~\eqref{eq:gx} can also be interpreted as the selection gradient on a continuously varying mixed strategy $x$ (denoting the probability that action A is played),
evolving according to the traditional breeder's equation or the canonical equation of adaptive dynamics~\citep{Dieckmann1996},
so that the dynamics is of the form
\begin{equation*}
 \frac{\dd x}{\dd t} = v(x)g(x),
\end{equation*}
for some measure $v(x)$ of genetic variance~\citep{Kirkpatrick2005}.
Hence, all our subsequent results pertaining to polymorphic equilibria in pure strategies can also be interpreted in terms of monomorphic equilibria for mixed strategies.
\end{rmk}

\section{Sign patterns and (the stability of) rest points}
\label{sec:pattern}

The fact that the gain function is a polynomial in Bernstein form implies a tight link between
the sign pattern of the gain sequence on the one hand
and the sign pattern and number of roots of the gain function on the other hand.
This is due to two properties of Bernstein polynomials,
namely the preservation of initial and final signs and the variation diminishing property (see Properties~\ref{lem:initial} and~\ref{lem:variation} below).
Because roots of the gain function correspond to interior rest points of the replicator dynamics and the sign pattern of the gain function informs us about the direction of selection at the trivial rest points as well as changes in the direction of selection at interior rest points, general results about the number and stability of rest points follow immediately (see Results \ref{lem:stable-triv} and \ref{prp:interior}). These results hold for any non-zero gain sequence, allow for interior rest points at which the direction of selection does not change, and provide more detailed information about the number of rest points and stable equilibria than the observations made by~\citet{Cuesta2007} and~\cite{Gokhale2010}. Results \ref{cor:simple} to \ref{cor:unimodal}  summarize the implications of the general results for gain sequences with at most two sign changes, providing the basis for our subsequent analysis.

\subsection{Preliminaries}
\label{sec:preliminaries}

To proceed, we require some terminology and notation to describe sign patterns \citep[see][]{Brown1981} and other relevant shape properties of gain sequences and gain functions. The same notation and terminology applies to other sequences and functions we encounter in our analysis.

Let $I(\dkk)$ denote the sign (either $+$ or $-$) of the first non-zero entry in the sequence $\dkk$.
Likewise, let $F(\dkk)$ denote the sign of the last non-zero entry in $\dkk$.
We refer to $I(\dkk)$ and $F(\dkk)$ as the initial and final signs of the gain sequence $\dkk$.
We also denote by $S(\dkk)$ the number of sign changes between consecutive entries in $\dkk$ after zero entries have been eliminated.
Obviously, $0 \leq S(\dkk) \leq n$.

As we have assumed $\dkk \not= \bm 0$, there exists a neighborhood of $x^*=0$ such that the sign of $g(x)$ is either $+$ or $-$ for all $x \not= 0$ in this neighborhood.
We define the initial sign $I(g)$ of $g(x)$ as the sign of $g(x)$ in such neighborhood, and define the final sign $F(g)$ in an analogous way.
Note that $I(g)$ coincides with the sign of $g(0)$  if $g(0) \not= 0$ holds. Similarly, if $g(1) \not= 0$ holds, then $F(g)$ coincides with the sign of $g(1)$.
The number of sign changes $S(g)$ of the function $g(x)$ in the interval $(0,1)$ is the number of times it crosses the $x$-axis in $(0,1)$.

The notation $\Delta \dkk = (\Delta d_0, \ldots, \Delta d_{n-1})$,
where $\Delta d_k \equiv d_{k+1} - d_k$,
denotes the (first) forward difference of the sequence $\dkk$.
The second forward difference of the sequence $\dkk$ is $\Delta^2 \dkk = (\Delta^2 d_0, \ldots, \Delta^2 d_{n-2})$, where $\Delta^2 d_k \equiv \Delta d_{k+1} - \Delta d_k$.
These forward differences can be viewed as the counterparts to the first and second derivatives of a real function and are a useful tool for describing the shape of a sequence.
In particular, the sequence $\dkk$ is increasing (resp.~decreasing) if $\Delta \dkk \ge \bm 0$ ($\Delta \dkk \le \bm 0$) holds,
convex (resp.~concave) if $\Delta^2 \dkk \ge \bm 0$ (resp.~$\Delta^2 \dkk \le \bm 0$) holds,
and unimodal (resp.~anti-unimodal) if the sequence $\Delta \dkk$ has a single sign change from positive to negative (resp.~from negative to positive).
\emph{Mutatis mutandis} the same definitions apply to the gain function $g(x)$. For instance, a gain function is unimodal if its first derivative $g'(x)$ has one sign change from positive to negative and is concave if its second derivative satisfies $g''(x) \le 0$ for all $0 \le x \le 1$.

\subsection{Stability of trivial rest points}

One important 
property of the Bernstein operator $\Bn_n$ is that it preserves end-points, i.e. $g(0) = \Bn_n(0;\dkk) = d_0$ and $g(1) = \Bn_n(1;\dkk) = d_n$~\citep{Farouki2012}.
From this, it is immediate that the initial and final signs of $g(x)$ and $\dkk$ coincide in the case when $d_0 \not=0$ and $d_n \not=0$.
We show in Appendix~\ref{app:initial} that the same conclusion obtains in general,
so that we have the following property.

\begin{property}[Preservation of initial and final signs]
\label{lem:initial}
The initial and final signs of $g(x)$ and $\dkk$ coincide. That is,
\begin{equation*}
I(g) = I(\dkk) \text{ and } F(g) = F(\dkk).
\end{equation*}
\end{property}

The initial sign of $g(x)$ describes the direction of selection in a vicinity of the trivial rest point $x = 0$, so that the rest point $x=0$ is stable if and only if the initial sign of $g(x)$ is negative. Similarly, the rest point $x=1$ is stable if and only if the final sign of $g(x)$ is positive.
Hence, Property~\ref{lem:initial} implies that the initial and final signs of the gain sequence are all the information required to determine the stability of the trivial rest points.
This is explicitly stated in the following result.

\begin{result}[Stability of trivial rest points] \label{lem:stable-triv}
\begin{enumerate}
\item[]
\item The rest point $x=0$ is stable if and only if $ I(\dkk) = - $. 
\item The rest point $x=1$ is stable if and only if $ F(\dkk) = + $. 
\end{enumerate}
\end{result}

The first part of Result \ref{lem:stable-triv} asserts that strategy A is disadvantageous when rare if and only if the first non-zero element in the gain sequence is negative.
The second part is the assertion that strategy A is advantageous when common if and only if the last non-zero element in the gain sequence is positive.

\subsection{Number of (stable) interior rest points}

Let $R(g) \geq 0$ denote the number of roots of $g(x)$ in the interval $(0,1)$, counting roots according to their multiplicity.
The following is the variation diminishing property of Bernstein polynomials. 

\begin{property}[Variation diminishing property] \label{lem:variation}
\begin{enumerate}
\item[]
\item The number of roots of $g(x)$ on $(0,1)$ is equal to the number of sign changes of $\dkk$ or less by an even amount. That is,
\begin{equation} \label{eq:roots}
R(g) = S(\dkk) - 2 i, \text{ where } i \ge 0 \text{ is an integer}.
\end{equation}
\item The number of sign changes of $g(x)$ is equal to the number of sign changes of $\dkk$ or less by an even amount. That is,
\begin{equation} \label{eq:signs}
S(g) = S(\dkk) - 2 j, \text{ where } j \ge i \text{ is an integer}.
\end{equation}
\end{enumerate}
\end{property}

The first part of the variation-diminishing property (see e.g.~\citet{Farouki2012}) follows from Descartes' rule of signs, which hence can be said to ``carry over'' to polynomials in Bernstein form. The second part follows from the first upon observing that $x \in (0,1)$ is the location of a sign change of $g(x)$ if and only if $x$ is a root of $g(x)$ with odd multiplicity, so that $S(g)$ is either equal to $R(g)$ or less by an even amount.

As the interior rest points of the replicator dynamics coincide with the roots of $g(x)$, Property \ref{lem:variation}.1 applies as stated to the interior rest points of the replicator dynamics. In particular, as noted by~\citet{Cuesta2007} and~\citet{Gokhale2010}, the number of sign changes of the gain sequence $\dkk$ provides an upper bound on the number of interior rest points. If the number of sign changes of $\dkk$ is odd,~\eqref{eq:roots} implies that $R(g)$ is odd. Consequently, the replicator dynamics possesses at least one interior rest point in this case.

Stability of an interior rest point is equivalent to the requirement that the sign of $g(x)$ changes from $+$ to $-$ at the rest point.
As sign changes must alternate and
initial signs are preserved (Property~\ref{lem:initial}), the second part of the variation diminishing property yields the following result.

\begin{result}[Number of stable interior rest points] \label{prp:interior}
Let $\ell$ denote the number of stable interior rest points of the replicator dynamics and let $j \ge 0$ be the integer appearing in the statement of Property \ref{lem:variation}.2.
\begin{enumerate}
\item If $S(\dkk)$ is even, then $\ell = S(g)/2 = S(\dkk)/2 -j$.
\item If $S(\dkk)$ is odd and $I(\dkk) = -$, then $\ell = (S(g)-1)/2 = (S(\dkk)-1)/2 - j$.
\item If $S(\dkk)$ is odd and $I(\dkk)=+$, then $1 \le \ell =(S(g)+1)/2 = (S(\dkk)+1)/2-j$.
\end{enumerate}
\end{result}

\subsection{Special cases}

It will be convenient to summarize the relationship between the sign patterns of the gain sequence
and the rest points of the replicator dynamics for the cases in which the gain sequence has at most two sign changes.
We also provide simple sufficient conditions ensuring that a gain sequence has at most one, resp.~two sign changes.

\subsubsection{Gain sequences with one or no sign change}
\label{sec:one}

When the gain sequence has no or one sign change, the variation diminishing property implies that the number of roots and the number of sign changes of the gain function both coincide with the number of sign changes of the gain sequence. In particular, Result \ref{prp:interior} holds with $j=0$. Combining these observations with Result \ref{lem:stable-triv} then shows that
for games with gain sequences having at most one sign change, the sign pattern of the gain sequence contains all the information required to determine the number and stability of rest points. For later reference we state the ensuing case distinction in the following result.

\begin{result}[Gain sequences with no or one sign change]
\label{cor:simple} 
\begin{enumerate}
\item[]
\item If the gain sequence has no sign changes, then the replicator dynamics has no interior rest points. Moreover
\begin{enumerate}
\item If $I(\dkk) = -$, then $x=0$ is stable and $x=1$ is unstable.
\item If $I(\dkk) = +$, then $x=0$ is unstable and $x=1$ is stable.
\end{enumerate}
\item If the gain sequence has a single sign change, then the replicator dynamics has a unique interior rest point $x^*$.
 Moreover:
 \begin{enumerate}
  \item If $I(\dkk)=-$, then $x=0$ and $x=1$ are stable, and $x^*$ is unstable.
  \item If $I(\dkk)=+$, then $x=0$ and $x=1$ are unstable, and $x^*$ is stable.
 \end{enumerate}
\end{enumerate}
\end{result}

 The four possible dynamical regimes appearing in Result \ref{cor:simple} correspond to the cases that are familiar from the evolutionary analysis of symmetric two-player games with two pure strategies \citep[see, e.g.][Section 2.2]{Cressman2003}. This is, of course, not a coincidence: such two-player games are nothing but the special case of our model with $n=1$ and thus feature gain sequences with at most one sign change.

A simple sufficient condition for the applicability of Result \ref{cor:simple} is that the gain sequence is monotonic, that is, either increasing or decreasing.
It is clear that an increasing gain sequence can have at most one sign change and that such a sign change occurs if and only if $d_0 < 0 < d_n$.
In this case, the rest points of the replicator dynamics are characterized by Result \ref{cor:simple}.2.a.
The other two possibilities for an increasing gain sequence, namely $d_n \le 0$ and $d_0 \ge 0$, are covered by Result \ref{cor:simple}.1.a and Result \ref{cor:simple}.1.b, respectively. Similarly, for a decreasing gain sequence only three of the four scenarios described in Result \ref{cor:simple} are possible, with a stable interior rest point occurring if and only if $d_0 > 0 > d_n$.

\subsubsection{Gain sequences with two sign changes}

If the gain sequence has two sign changes, its initial and final signs coincide.
Suppose they are both negative.
Then, by the preservation of initial and final signs (Property \ref{lem:initial}), the same is true for the initial and final signs of $g(x)$.
In particular, as indicated by Result \ref{lem:stable-triv}, the rest point $x=0$ is stable and the rest point $x=1$ is unstable.
Further, the first part of the variation diminishing property implies that the replicator dynamics has either
(i) two distinct interior rest points (which correspond to simple roots in which $g(x)$ crosses zero),
(ii) one interior rest point (corresponding to a double root in which $g(x)$ touches, but does not cross zero),
or (iii) no interior rest point. In the first of these cases $g(x)$ has two sign changes and the larger of the two interior rest points is stable. In the other two cases $g(x)$ has no sign change and, consequently, no stable interior rest point. Considering the maximal value of $g(x)$ on $[0,1]$, which we denote by $\bar g$, provides a convenient way to describe which of these three cases arises. In particular, for $\bar g < 0$ there is no interior rest point, for $\bar g = 0$ there is exactly one interior rest point, and for $\bar g > 0$ there are two interior rest points.
Analogous reasoning can be applied for the case in which the initial and final signs are both positive.
These considerations are summarized in the following result.

\begin{result}[Gain sequences with two sign changes]
 \label{prp:dktwosignchanges}
 Let $\bar g = \max_{0 \le x \le 1} g(x)$ and $\underline g = \min_{0 \le x \le 1} g(x)$. Then:
 \begin{enumerate}
    \item If $S(\dkk) = 2$ and $I(\dkk)=-$ the rest point $x=0$ is stable and the rest point $x=1$ is unstable. Further:
  \begin{enumerate}
   \item if $\bar g < 0$, the replicator dynamics has no interior rest points.
   \item if $\bar g= 0$, then the replicator dynamics has one interior rest point $\hat x$ which is unstable.
   \item if $\bar g > 0$, the replicator dynamics has one unstable rest point $\xl$ and one stable rest point $\xr$, satisfying $0 < \xl < \xr < 1$.
  \end{enumerate}
  \item If $S(\dkk) = 2$ and $I(\dkk)=+$ the rest point $x=0$ is unstable and the rest point $x=1$ is stable. Further:
  \begin{enumerate}
   \item If $\underline g > 0$, the replicator dynamics has no interior rest points.
   \item If $\underline g = 0$, the replicator dynamics has one interior rest point $\hat{x}$ which is unstable.
   \item If $\underline g < 0$, the replicator dynamics has one stable rest point $\xl$ and one unstable rest point $\xr$, satisfying $0 < \xl < \xr < 1$.
  \end{enumerate}
 \end{enumerate}
\end{result}

It is evident from the case distinctions appearing in Result \ref{prp:dktwosignchanges} that for gain sequences with two sign changes, information beyond the one contained in the sign pattern of the gain sequence is required to determine the number of interior rest points. However, the additional information required takes a simple form (namely, the knowledge of the maximal, resp. minimal value of the gain function),
which is amenable to further analysis.

\begin{rmk}
If a gain sequence has more than two sign changes, Results \ref{lem:stable-triv} and \ref{prp:interior} still provide useful information about the possible range of dynamical scenarios,
but determining which of these scenarios arises becomes much harder than in the case of at most two sign changes.
To illustrate this, consider the case $S(\dkk) = 3$ and suppose $I(\dkk) = +$. We then have $F(\dkk) = -$, implying that both trivial rest points are unstable (Result \ref{lem:stable-triv}). Furthermore, there are either one or two stable interior rest points (Result \ref{prp:interior}). In the second of these cases there must exist a single unstable interior rest point, in the first case there is either no unstable interior rest point or one unstable interior rest point which corresponds to a root of the gain function with multiplicity two (Property \ref{lem:variation}.1).
\end{rmk}

\subsubsection{Unimodal gain sequences}
\label{sec:unimodal}

Unimodality or anti-unimodality is a simple sufficient condition ensuring that a gain sequence has at most two sign changes. Furthermore, a complete classification of the possible dynamic scenarios is easily obtained. Here we demonstrate these claims for the unimodal case; the argument (and result) for the anti-unimodal case is analogous.

Our argument relies on the identity
\begin{equation} \label{eq:deriv}
g'(x) = n \Bn_{n-1}(x; \Delta \dkk),
\end{equation}
which is a classical result in approximation theory, known as the derivative property of polynomials in Bernstein form \citep[see e.g.][]{Lorentz1986,DeVore1993,Farouki2012}.
By~\eqref{eq:deriv} the derivative $g'(x)$ is proportional to a Bernstein polynomial with coefficients $\Delta \dkk$.
We may thus apply Properties \ref{lem:initial} and \ref{lem:variation} to the relationship between the sign pattern of $\Delta \dkk$ and the roots and sign pattern of $g'(x)$.
Recalling that for a unimodal gain sequence $\Delta \dkk$ has a single sign change from positive to negative, it follows that unimodality of the gain sequence implies unimodality of the gain function. Moreover, applying the first part of the variation diminishing property, there exists a unique $0 < \hat x < 1$ satisfying the first order condition $g'(\hat x) = 0$. Unimodality of $g(x)$ implies that $\hat x$ is the unique solution to the problem $\max_{0 \le x \le 1} g(x)$ appearing in the statement of Result \ref{prp:dktwosignchanges}. In particular, we have $\bar g = g(\hat x)$.

It is clear that a unimodal gain function can have at most one sign change in its increasing part (which then must be from negative to positive)
and at most one sign change in its decreasing part (which then must be from positive to negative).
Moreover, a sign change in the increasing part occurs if and only if $g(0) < 0 < g(\hat x)$ and a sign change in the decreasing part occurs if and only if $g(1) < 0 < g(\hat x)$.
Combining these observations yields the following result, refining Results \ref{cor:simple} and \ref{prp:dktwosignchanges} for the unimodal case.

 \begin{result}[Unimodal gain sequences]
 \label{cor:unimodal}
 If the gain sequence is unimodal, there exists a unique $0 < \hat x < 1$ solving the equation $g'(\hat x) = 0$. Moreover:
 \begin{enumerate}

    \item If $g(\hat x) < 0$, then the replicator dynamics has no interior rest point. The rest point $x=0$ is stable and the rest point $x=1$ is unstable.
    \item If $g(\hat x) = 0$, then $\hat x$ is the unique interior rest point of the replicator dynamics. The rest point $x=0$ is stable and the rest points $\hat x$ and $x=1$ are unstable.
    \item If $g(\hat x) > 0$ holds, then one of the following four cases applies:
        \begin{enumerate}
    \item If $\min \{d_0, d_n \} \ge 0$, then the replicator dynamics has no interior rest point. The rest point $x=0$ is unstable and the rest point $x=1$ is stable.
    \item If $\max \{d_0, d_n\} < 0$, then the replicator dynamics has two interior rest points satisfying $\xl < \hat x < \xr$. The rest points $x=0$ and $\xr$ are stable, whereas the rest points $\xl$ and $x=1$ are unstable.
    \item If $d_0 < 0$ and $d_n \ge 0$, then the replicator dynamics has a unique interior rest point $x^* < \hat x$. The rest points $x=0$ and $x=1$ are stable, whereas the rest point $x^*$ is unstable.
     \item If $d_0 \ge 0$ and $d_n < 0$, then the replicator dynamics has a unique interior rest point $x^* > \hat x$. The rest point $x^*$ is stable, whereas the rest points $x=0$ and $x=1$ are unstable.
         \end{enumerate}
   \end{enumerate}
\end{result}

\begin{rmk}
Using the derivative property of polynomials in Bernstein form, it can be shown that all the properties of gain sequences mentioned at the end of Section~\ref{sec:preliminaries} are inherited by the gain function
(e.g., if the gain sequence is increasing, so is the gain function).
The argument for the preservation of anti-unimodality is analogous to the one we have given for the preservation of unimodality.
The other results are well known properties of Bernstein polynomials, namely preservation of monotonicity, and~preservation of convexity \citep[see][]{Lorentz1986,Farouki2012}. Seemingly unaware of these properties, \cite{Motro1991} proves preservation of monotonicity and \cite{Bach2006} prove preservation of concavity (which is equivalent to preservation of convexity).
\end{rmk}

\section{Public goods games}
\label{sec:pggs}

In this section, we apply Results \ref{cor:simple} to \ref{cor:unimodal} to two classes of public goods games, subsuming many of the models
encountered in the literature of the evolution of cooperation and collective action.

\subsection{Gain sequences for public goods games} \label{sec:gain-pgg}

In a \emph{public goods game}, playing A means to cooperate (i.e.~to contribute to the creation or maintenance of a public good)
and playing B means to defect (i.e.~to free ride on the contributions of others).
Contributing entails a cost $c_k \ge 0$ to the focal cooperator, where $k$ is the number of other cooperators. Defectors bear no cost.
All players obtain a benefit $r_j \ge 0$ from the public good, where $j$ is the total number of cooperators in the group. Note that for a focal cooperator $j=k+1$, while for a focal defector $j=k$. With these assumptions, the payoff sequences for a public goods game can thus be written as
\[
 a_k = r_{k+1} - c_k ,  \ k=0,1,\ldots, n
\]
and
\[
 b_k = r_k , \ k=0,1,\ldots, n
\]
so that the gain sequence is given by
\begin{equation}
\label{eq:dkpgg}
 d_k = \Delta r_{k} - c_k , \ k=0,1,\ldots, n.
\end{equation}

In all public goods games we consider the benefit sequence $\rkk = \rk$ is increasing and neither the cost sequence $\ckk = \ck$ nor the first forward difference of the benefit sequence $\Delta \rkk$ are equal to zero.

If no further assumptions are imposed on the cost and benefit sequence, it is clear from~\eqref{eq:dkpgg} that any $\dkk$ can arise as the gain sequence of a public goods game.
Consequently, to obtain insights into the evolutionary dynamics of public goods games going beyond the ones summarized in Results \ref{lem:stable-triv} and \ref{prp:interior}, additional assumptions on the benefit or the cost sequence are required.
In this light, it is not surprising that the public goods games usually studied in the biological literature fall into one of the two classes that we discuss in the following subsections.

\subsection{Threshold games}

If there exists an integer $m$ with $1 \le m \le n+1$ and a constant $r > 0$ such that the benefit sequence satisfies  $r_j = 0$ if $j < m$ and $r_j = r$ if $j \geq m$,
we say that a public goods game is a \emph{thres\-hold game}.
This class of games describes situations in which the public good is a ``step good'' in the sense of \citet[p. 55]{Hardin1982}:
at least $m$ cooperators are required to provide a public good for all group members,
but the number of cooperators beyond the threshold $m$ does not increase the benefit received by the players.
Examples of such threshold games abound in the theoretical literature of the social sciences~\citep{Hardin1982,Taylor1982,Diekmann1985,Sugden1986,Weesie1998,Hoffler99,Herold2012}
and evolutionary biology~\citep{Dugatkin1990,Bach2006,Zheng2007,Archetti2009,Souza2009},
and are sometimes referred to as volunteer's dilemmas or multi-player snowdrift games.

For threshold games \eqref{eq:dkpgg} reduces to
\begin{equation}
\label{eq:thres}
 d_k =
 \left\{
   \begin{array}{ll}
     - c_k               & \mbox{if } k < m -1 \\
     r - c_{m-1} & \mbox{if } k = m -1\\
     - c_k               & \mbox{if } k > m -1
     \end{array}
 \right. .
\end{equation}
It is obvious that the gain sequence $\dkk$ has no sign change when $r \le c_{m-1}$ and that in this case defection is a dominant strategy.
As illustrated in Fig.~\ref{fig:threshold} and discussed below, in the other cases the sign pattern of the gain sequence depends on the location of the threshold $m$.

\subsubsection{Threshold $m=1$}
\label{sec:thres-1}

Threshold games with $m = 1$ represent situations in which only one cooperator is required for the provision of the public good.
Such games have been considered by~\cite{Dugatkin1990},~\cite{Weesie1998},~\cite{Zheng2007}, and~\cite{Souza2009}
for the particular case of a cost sequence satisfying $c_k = c/(k+1)$ for some constant $c > 0$,
so that the cost to cooperators is inversely proportional to the total number of cooperators in the group.
These authors have shown by algebraic manipulations or numerical simulations that for such games the replicator dynamics has at most one interior stable rest point. \citet{Archetti2009} shows the same result for a cost sequence satisfying $c_k = c$ for some constant $c > 0$.

Considering the sign pattern of the gains from switching not only recovers this result in a simpler way, but also extends it to any strictly positive cost sequence $\ckk$.
If $r > c_0$, the gain sequence given in~\eqref{eq:thres} has exactly one sign change and $I(\dkk) = +$,
so that Result~\ref{cor:simple}.2.b establishes the existence of a single interior stable rest point $0 < x^* < 1$ and the instability of the trivial rest points (see Fig.~\ref{fig:threshold}.a).
If $r \le c_0$, Result \ref{cor:simple}.1.a applies. Hence, there is no interior rest point and  $x= 0$ is the unique stable rest point.

\subsubsection{Threshold $m=n+1$}
\label{sec:thres-2}

Recalling that $N = n+1$ is group size,
threshold games with $m=n+1$ represent situations in which the cooperation of all group members is required to produce the public good.
For the case $m=n+1=2$ and a cost sequence satisfying $0 < c_0 = c_1 < r$,~\cite{Souza2009} observe that such a thres\-hold game corresponds to a two-player stag hunt game~\citep{Skyrms2004}
in which both trivial rest points are stable and there is a unique, unstable interior rest point. It is easy to see that this result holds more generally. Indeed, provided that the cost sequence is strictly positive and satisfies $r > c_n$,
the gain sequence given in~\eqref{eq:thres} is characterized by $S(\dkk)=1$ and $I(\dkk) = -$.
Then, by Result~\ref{cor:simple}.2.a, it follows that
the qualitative dynamics of the two-player stag hunt
are recovered for every threshold game with $m = n+1$ (see Fig.~\ref{fig:threshold}.b).
The case $r \le c_n$ is covered by Result~\ref{cor:simple}.1.a.

\subsubsection{Threshold $1 < m < n+1$}
\label{sec:thres-int}

\cite{Souza2009} studied a threshold game with $1 < m < n+1$ for a cost sequence of the form
 \begin{equation}
 \label{eq:cost_thres}
 c_k =
 \left\{
   \begin{array}{ll}
     c/m                             & \mbox{if } k < m-1 \\
     c/(k+1)                         & \mbox{if } k \geq m-1
 	\end{array}
 \right.
\end{equation}
for some constant $c > 0$. Their main theoretical result~\citep[Theorem 1]{Souza2009} uses an ingenious but rather involved argument to demonstrate that in this example there exists $\bar c > 0$ and $0 < \bar x < 1$ such that
(i) if $c < \bar c$, the replicator dynamics has two interior rest points $\xl < \bar x < \xr$ where $\xl$ is unstable and $\xr$ is stable (see Fig.~\ref{fig:threshold}.c),
(ii) if $c = \bar c$, the replicator dynamics has a unique rest point $\bar x$ (which is unstable), and
(iii) if $c > \bar c$, the replicator dynamics has no interior rest point (see Fig.~\ref{fig:threshold}.d).\footnote{\citet{Souza2009} express their results in terms of the cost-benefit ratio $c/r$. The difference is of no importance as time can always be rescaled to ensure $r = 1$.}

In Appendix~\ref{app:souza} we prove that the same result holds for any cost sequence of the form $c_k = c \cdot \gamma_k$, where the strictly positive, but otherwise arbitrary, sequence $\bm{\gamma}$ describes the shape of the cost sequence and, as in the example considered by \citet{Souza2009}, $c$ shifts the level of the cost sequence.
Our result follows, in essence, from two observations.
The first is that for every threshold game with $1 < m < n+1$ and strictly positive cost sequence satisfying $0 < c_{m-1} < r$ the gain sequence has two sign changes and a negative initial sign, so that the rest points of the replicator dynamics are described by Result~\ref{prp:dktwosignchanges}.1.
The second observation is that the maximal value of the gain function $\bar g$ is strictly decreasing in the cost parameter $c$.

Threshold games with $1 < m < n+1$ have also been considered by \citet{Bach2006}, \citet{Archetti2009}, and \citet{Archetti2011}. These authors assume a cost sequence satisfying $c_k = c$ for some constant $c > 0$,  implying that these games fall in the class of constant cost games with sigmoid benefit functions that we discuss in Section~\ref{sec:sigmoid}.

\subsubsection{Further threshold games}

In economics,~\citet{Hoffler99} and~\citet{Herold2012} have studied evolutionary dynamics of threshold games which differ from the biological threshold games considered above in that cooperators pay a cost only if the threshold for the successful provision of the public good is reached. In such cases the gain sequence has the form
\begin{equation}
\label{eq:thres-econ}
 d_k =
 \left\{
   \begin{array}{ll}
     0               & \mbox{if } k < m -1 \\
     r - c_{m-1} & \mbox{if } k = m -1\\
     - c_k               & \mbox{if } k > m -1
     \end{array}
 \right.
\end{equation}
and thus possesses at most one sign change (see Fig.~\ref{fig:hoffler}).
For $r > c_{m-1}$ and $1 \leq m < n+1$, this gain sequence satisfies $I(\dkk) = +$ and $S(\dkk) = 1$.
Applying Result~\ref{cor:simple}.2.b then yields a simple direct proof of the main result obtained by \citet[Proposition 1]{Hoffler99} and \citet[Proposition 1]{Herold2012} for this class of games, namely that there exists a unique stable interior rest point.\footnote{Proposition 2 in \citet{Hoffler99}, which considers the case $m = n+1$, is implied by our Result~\ref{cor:simple}.1.b. Herold also considers the case in which cooperators only pay a cost if the threshold is \emph{not} reached. His main result for this case \citep[Proposition 2]{Herold2012} is implied by our Result~\ref{cor:simple}.2.a.}

\subsection{Constant cost games}

If there exists a constant $c > 0$ such that $c_k = c$ holds for $k=0,\ldots,n$ we say that a public goods game is a \emph{constant cost game}.
Such games have been studied, among others, by \citet{Motro1991},~\citet{Szathmary1993},~\citet{Bach2006},~\citet{Hauert2006a},~\citet{Pacheco2009}, and~\citet{Archetti2011}.

In the case of a constant cost game, equation~\eqref{eq:dkpgg} reduces to
\begin{equation} \label{eq:cocost}
d_k = \Delta r_{k} - c, \ k=0,1,\ldots, n.
\end{equation}
It is then immediate that the gain sequence has no sign change (and hence no interior rest point) if $c \ge \max_{k=0, \ldots, n} \Delta r_k$ or $\min_{k=0, \ldots, n} \Delta r_k \ge c$ holds. It follows from Result~\ref{cor:simple}.1 that in the former case $x=0$ and in the latter case $x=1$ is the unique stable rest point. In all other cases, that is whenever the inequality
\begin{equation} \label{eq:coco-int}
\min_{k=0, \ldots, n} \Delta r_k < c < \max_{k=0, \ldots, n} \Delta r_k
\end{equation}
holds, the gain sequence has at least one sign change.

In the following, we consider three different kinds of constant cost games, arising from three different assumptions on the shape of the benefit sequence:
linear benefits (Section~\ref{sec:linpgg}), convex or concave benefits (Section~\ref{sec:con_benefits}) and sigmoid benefits (Section~\ref{sec:sigmoid}).
See Fig.~\ref{fig:pggs} for a graphical illustration of these different constant cost games.

\subsubsection{Linear benefits}
\label{sec:linpgg}

The familiar linear public goods game
is a constant cost game in which the benefit sequence is given by $r_j = j r/(n+1)$~\citep{Sigmund2010}. The interpretation is that $r > 0$ is the amount of the public good produced by each cooperator and that this amount is split evenly among the $N = n+1$ members of the group. For such a game, we have $\Delta r_k = r/(n+1)$, so that the gain sequence is $d_k = r/(n+1) - c$, which is a constant independent of $k$. Hence $\dkk$ has no sign change. Making the standard assumption $r < (n+1) c$, we have $I(\dkk) = -$, so that there are no interior rest points and $x=0$ is the unique stable rest point (see Fig.~\ref{fig:pggs}.a).
This conclusion is, of course, well-known.

\subsubsection{Convex or concave benefits}
\label{sec:con_benefits}

Convexity of the benefit sequence ($\Delta^2 \bm r \ge \bm{0}$)
indicates that the incremental benefit $\Delta r_k$ of a further contributor is increasing in the number of other contributors $k$ that are already present in the group.
Using \eqref{eq:cocost} to obtain
\begin{equation} \label{eq:second-pgg}
\Delta d_k = \Delta^2 r_k, \ k=0,1,\ldots, n-1,
\end{equation}
it is apparent that that the gain sequence $\dkk$ is increasing. As discussed in Section~\ref{sec:one} it follows that~\eqref{eq:coco-int} reduces to $\Delta r_0 < c < \Delta r_n$. Furthermore, if these inequalities hold, Result \ref{cor:simple}.2.a implies that there is a unique interior rest point which is unstable, whereas both trivial rest points are stable (see Fig.~\ref{fig:pggs}.b). Similarly, when the benefit sequence is concave ($\Delta^2 \bm r \le \bm{0}$),~\eqref{eq:coco-int} reduces to $\Delta r_n < c < \Delta r_0$ and if these inequalities hold, Result \ref{cor:simple}.2.b  implies there is a unique interior rest point which is stable, whereas both trivial rest points are unstable (see Fig.~\ref{fig:pggs}.c).

The argument we have just given recovers the main results from~\citet{Motro1991}.
A simple illustration of a constant cost game with convex or constant benefits
is provided by the model of synergy and discounting considered in \citet[Section 2.1]{Hauert2006a}. These authors consider a constant cost game with benefit function
\begin{equation} \label{eq:hauert-1}
 r_j = \frac{r}{n+1}\left(1 + w + \ldots w^{j-1}\right),
\end{equation}
where $r > 0$ and $w > 0$ are parameters. For this specification we have $\Delta r_k = r w^k/(n+1)$.
For $w > 1$ this benefit sequence is convex, whereas for $w < 1$ it is concave. The case $w=1$ is the linear public goods game.
We observe that the classification obtained in Section 2.2 of~\citet{Hauert2006a},
corresponds to the one obtained from a straightforward application of our Result~\ref{cor:simple}. 

\subsubsection{Sigmoid benefits}
\label{sec:sigmoid}

A benefit sequence is sigmoid (or S-shaped) when $\Delta^2 \bm r$ has exactly one sign change from $+$ to $-$, i.e. the benefit sequence is first convex, then concave.
Examples of sigmoid benefit sequences are the threshold benefit sequences with $1 < m < n+1$ considered in Section \ref{sec:thres-int},
the ``benefit function with a hump'' proposed in~\cite{Szathmary1993}, and the threshold-linear and logistic benefit sequences
studied respectively by \cite{Pacheco2009} and \citet{Archetti2011}.

In this case it is immediate from~\eqref{eq:second-pgg}
that the gain sequence of a constant cost game with sigmoid benefits is unimodal.
Consequently, the characterization of the different types of dynamics that can arise in such games
involves nothing more than inserting the values $d_k = \Delta r_k - c$ into our Result~\ref{cor:unimodal} (see Fig.~\ref{fig:pggs}.d for a particular example).
The results of this exercise have been published by~\citet{Archetti2013}.%
\footnote{\citet{Archetti2013} ignores most of the cases in which a weak inequality occurs in Result~\ref{cor:unimodal} and neglects to impose the proper sign change condition required for unimodality, but these shortcomings are easily fixed.}

Sigmoid benefit sequences generalize the benefit sequences considered in
\citet[Proposition 7]{Bach2006}, who not only assume that $\Delta^2 \rkk$ has a single sign change from $+$ to $-$, but, in addition, require $\Delta^2 \rkk$ to be decreasing.
Using these assumptions, ~\citet{Bach2006} establish the existence of a $c^* > \max \{\Delta r_0, \Delta r_n\}$ such that for $c < c^*$ the replicator dynamics has two interior rest points (the larger of which is stable), whereas for $c = c^*$ there is a unique (unstable) interior rest point and for $c > c^*$ there is none.
As the gain sequence (and hence the gain function and $\bar g$) for constant cost games is linearly decreasing in $c$, it is immediate from Result~\ref{cor:unimodal} that the same conclusion obtains for all sigmoid benefit sequences.

\section{Other multi-player games}

Up to this point our examples have considered public goods games. 
Here we consider two examples of other multi-player games,
illustrating how focusing on the shape of the gain sequence obviates the need for a more involved analysis.
Of course, further examples could be analyzed along similar lines.
For instance, it is straightforward to show that in the ``shared reward dilemma" considered by~\citet{Cuesta2008}
the gain sequence has at most two sign changes, so that we can recover their case distinctions by applying our results.

\subsection{Repeated $N$-person prisoner's dilemma}

\citet{Joshi1987}, \citet{Boyd1988} and \citet{VanSegbroeck2012} considered a repeated $N$-person prisoner's dilemma with two possible strategies.
Reciprocators (A-strategists) contribute to the public good in the first round and then contribute in each subsequent round if at least $m$ individuals (including the focal individual)
contributed in the previous move.
Defectors (B-strategists) never contribute to the public good. Payoffs in each round depend on the number of contributors as in the linear public goods game considered in Section \ref{sec:linpgg}.

The gain sequence for this model is easily derived by considering the first round and the subsequent rounds separately.
In the first round, the gain if switching from B to A is $r/(n+1) - c < 0$.
In each subsequent round, the gain from switching is zero if $k < m-1$ (because all players defect),
$r/(n+1)-c$ if $k > m-1$ (because the other reciprocators cooperate no matter whether the focal individual contributes or not), and $m r/(n+1) - c$ if $k = m-1$ (because in this case the contribution of the focal individual in the first round is pivotal in determining the subsequent behavior of reciprocators). Setting
\[
\tilde c = c - r/(n+1) > 0 ,
\]
and
\[
\tilde r = (m-1)r/(n+1) ,
\]
the gain sequence can be written as
\begin{equation}
\label{eq:dkrnpd}
d_k =
 \left\{
   \begin{array}{ll}
         - \tilde c                                   & \mbox{if } k < m-1 \\
         T \tilde r  - (T+1) \tilde c  & \mbox{if } k = m-1 \\
         - (T+1) \tilde c                          & \mbox{if } k > m-1
     \end{array}
 \right. ,
\end{equation}
where $T > 0$ denotes the expected number of rounds after the first one.
From~\eqref{eq:thres} and~\eqref{eq:dkrnpd} it is apparent that the model is equivalent to a threshold game with 
(i) the benefit $T \tilde r$ arising if and only if at least $m$ reciprocators are present and
(ii) costs given by $c_k = \tilde c$ if $k < m-1$ and $c_k = (T+1) \tilde c$ otherwise. In particular, the results for the cases $m=1$ and $m= n+1$ are identical to the ones discussed in Sections \ref{sec:thres-1} and \ref{sec:thres-2}. Moreover, when $T \tilde r - (T+1) \tilde c$ is negative, it is immediate that the gain sequence is negative and Result~\ref{cor:simple}.1.a applies.

In the remaining case, satisfying $1 < m < n+1$ and $T \tilde r - (T+1) \tilde c > 0$,
it follows from \eqref{eq:dkrnpd} that the only non-zero elements of $\Delta \bm{d}$ are $\Delta d_{m-2} > 0$ and $\Delta d_{m-1} < 0$.
Consequently, the gain sequence is unimodal and Result~\ref{cor:unimodal} applies with $\max \{d_0, d_n \} < 0$ to characterize the three different possible dynamical regimes.
Which of these regimes arises depends on the value of $\bar g = g(\hat x)$ (see Fig.~\ref{fig:repeatednpd} for an example of the case $\bar g > 0$).
As in all applications of Results \ref{prp:dktwosignchanges} and \ref{cor:unimodal}, a key question is whether this value can be linked to the parameters of the model.

For the parameter $T$ this question can be answered by using the linearity of the Bernstein operator $\Bn_n$ to write the gain function as
\begin{equation} \label{g-function}
g(x) = T h(x)  - \tilde c,
\end{equation}
where $h(x) = \Bn_n(x, \bm {e})$ and the sequence $\bm{e}$ is given by
  \[
\label{eq:dkrepeatednpd}
 e_k =
 \left\{
   \begin{array}{ll}
         0                                   & \mbox{if } k < m-1 \\
         \tilde r - \tilde c                        & \mbox{if } k = m-1 \\
         - \tilde c                          & \mbox{if } k > m-1
     \end{array}
 \right. .
\]
It follows from \eqref{g-function} that the critical value $\hat x$ satisfying the first order condition $g'(\hat x) = 0$ is independent of $T$. Further, because $I(\bm{e}) = +$, it follows from the preservation of initial signs that $h(\hat x) > 0$ holds. This in turn implies from \eqref{g-function} that $g(\hat x)$ is strictly increasing in $T$ and that the equation $\hat T = \tilde c/h(\hat x)$ identifies the critical value of $T$ at which $g(\hat x) = 0$ holds.
Hence, we obtain the same conclusions as \citet{VanSegbroeck2012} by an application of Result~\ref{cor:unimodal}.
Namely,
(i) for $T < \hat T$ there is no interior rest point,
(ii) for $T = \hat T$ the replicator dynamics has a single, unstable interior rest point,
and (iii) for $T > \hat T$ two interior rest points emerge.

\subsection{Constant cost game with different benefit sequences for cooperators and defectors}

\citet[Section 2.3.2]{Hauert2006a} consider an interesting extension of constant cost games
by allowing for the possibility that cooperators and defectors might obtain different benefits,
say $r^A_j$ and $r^B_j$, when there are $j$ cooperators in the group (see Fig.~\ref{fig:hauert}).
The counterpart to \eqref{eq:second-pgg}
is then
$\Delta d_k = \Delta r^A_{k+1} - \Delta r^B_k$. %
For the particular choice of benefit sequences in \citet{Hauert2006a}, given by \eqref{eq:hauert-1} for $r^A_j$ and
\begin{equation*}
 r^B_j = \frac{r}{n+1}\left(1 + v^1 + \ldots v^{j-1}\right),
\end{equation*}
this reduces to
\begin{equation} \label{eq:hauert-more}
\Delta d_k = \frac{r}{n+1} \left(w^{k+1} - v^k \right),
\end{equation}
where $r > 0$, $v > 0$ and $w > 0$ are parameters and $N = n+1$ is group size.

\cite{Hauert2006a} state that ``only $v=w$ allows for an analytical solution [...] but
in general there are [...] up to $N-1$ equilibria [rest points] in $(0,1)$."
Here we refine this statement and show that, as conjectured by~\citet{Cuesta2007}, the maximum number of interior rest points is two independently of group size.
To do so, we observe that $\Delta d_k > 0$ holds if and only if
\begin{equation*} 
 w > \left(\frac{v}{w}\right)^k .
\end{equation*}
Since the right side of this inequality is monotonic in $k$, equation \eqref{eq:hauert-more} implies the following, exhaustive case distinction:
\begin{enumerate}
\item if $w \ge 1$ and $w^{n} \ge v^{n-1}$ holds, then the gain sequence is increasing and there is at most one interior rest point (see Fig.~\ref{fig:hauert}.a).
\item if $w \le 1$ and $w^{n} \le v^{n-1}$ holds, then the gain sequence is decreasing and there is at most one interior rest point (see Fig.~\ref{fig:hauert}.b).
\item if $w > 1$ and $w^{n} < v^{n-1}$ holds, then the gain sequence is unimodal and there are at most two interior rest points (see Fig.~\ref{fig:hauert}.c).
\item if $w < 1$ and $w^{n} > v^{n-1}$ holds, then the gain sequence is anti-unimodal and there are at most two interior rest points (see Fig.~\ref{fig:hauert}.d).
\end{enumerate}

\section{Discussion}

Bernstein polynomials were first proposed more than a century ago by~\cite{Bernstein1912} in order to provide a constructive proof
of Weierstrass's approximation theorem~\citep{DeVore1993}. 
More recently, and because of their many shape-preserving properties, polynomials in Bernstein form have proven extremely useful in the field of computer aided geometric design~\citep{Yamaguchi1988,Farin2002}.
Here we have made the case for utilizing the shape-preserving properties of Bernstein polynomials in the analysis of multi-player matrix games.
In particular, we have used these properties to show how key insights into the evolutionary dynamics of multi-player matrix games can be obtained from studying the sign pattern of the gains from switching.

The properties of Bernstein polynomials we have used in this paper are certainly not the only ones of relevance for the theoretical analysis of collective action problems.
For instance, both the effects of changes in the group size~\citep[studied previously in][]{Motro1991} and the group size distribution~\citep[studied previously in][]{Pena2012} on the evolutionary dynamics can be analyzed by making use of the theory of polynomials in Bernstein form.
Our methods can also be extended to structured populations and used to analyze multi-player matrix games played between relatives.

\section*{Acknowledgements}

This work was supported by Swiss NSF grants PBLAP3-145860 (JP) and PP00P3-123344 (LL).

\newpage

\appendix

\titleformat{\section}{\normalfont\large\bfseries\scshape}{Appendix \thesection:}{1ex}{#1}
\numberwithin{equation}{section}

\section{Proof of Result \ref{lem:initial}} \label{app:initial}

We show the result $I(g) = I(\dkk)$; the argument that the final signs coincide is analogous. Using the derivative property of polynomials in Bernstein form (cf.~equation \eqref{eq:deriv}) recursively, for $0 \le m \le n$ the $m$-th derivative of the gain function can be written as~\citep{Farouki2012}
\begin{equation} \label{eq:higher-diff}
g^{(m)}(x) = n(n-1) \ldots (n-m+1) \Bn_{n-m}\left(x;\Delta^m \dkk \right),
\end{equation}
where (with the obvious iterative definition) $\Delta^m \dkk$ is the $m$-th forward difference of the sequence $\dkk$. Evaluating \eqref{eq:higher-diff} at $x= 0$ we obtain
\begin{equation} \label{eq:higher-diff-zero}
g^{(m)}(0) = n(n-1) \ldots (n-m+1) \Delta^m d_0.
\end{equation}
Now, let $\ell$ be the lowest index $k$ such that $d_{\ell} \not= 0$.
Then $\Delta^m d_0 = 0$ holds for all $m < \ell$ and $\Delta^{\ell} d_0 = d_{\ell}$.
Equation~\eqref{eq:higher-diff-zero} then implies that $g^{(m)}(0) = 0$ for all $m < \ell$
and that the sign of $g^{(\ell)}(0)$ coincides with the sign of $d_{\ell}$ which, by definition, is the initial sign of $\dkk$.
A standard Taylor-series argument as given in~\citet[Proof of Proposition 4]{Bach2006} demonstrates that the initial sign of $g$ coincides with the sign of $d_{\ell}$, finishing the proof.

\newpage

\section{Proof of the generalization of Theorem 1 from \citet{Souza2009}}
\label{app:souza}

For any $c \ge 0$ let
\begin{equation} \label{eq:gain-par}
g(x,c) = \sum_{k=0}^{n} \binom{n}{k} x^k (1-x)^{n-k} d_k(c),
\end{equation}
where
\begin{equation}
\label{eq:thres-par}
 d_k(c) =
 \left\{
   \begin{array}{ll}
     - c \gamma_k               & \mbox{if } k < m -1 \\
     r - c \gamma_{m-1} & \mbox{if } k = m -1\\
     - c \gamma_k               & \mbox{if } k > m -1
     \end{array}
 \right.
 \end{equation}
 and $\bm \gamma = (\gamma_0, \ldots, \gamma_n)$ is a given, strictly positive sequence. Let $\bar g(c) = \max_{0 \le x \le 1} g(x,c)$ denote the corresponding maximal value of the gain function.

For $0 < c < r/\gamma_{m-1}$ the gain sequence given in \eqref{eq:thres-par} satisfies $I(\dkk(c)) = -$ and $S(\dkk(c)) = 2$, so that the rest points of the replicator dynamics are described by Result~\ref{prp:dktwosignchanges}.1.

From \eqref{eq:gain-par} and \eqref{eq:thres-par} the function $g(x,c)$ is continuous. From the maximum theorem \citep[Theorem 9.14]{Sundaram1996} this ensures continuity of $\bar g(c)$. Because all the Bernstein coefficients $d_k(c)$ are strictly decreasing in $c$, every of the summands appearing in \eqref{eq:gain-par} is strictly decreasing in $c$, implying that $g(x,c)$ is strictly decreasing in $c$. This monotonicity property obviously carries over to $\bar g(c)$.

Consider the Bernstein coefficients as given in \eqref{eq:thres-par}. If $c = 0$, the only non-zero coefficient is $d_{m-1}(0) = r > 0$. It is then immediate from \eqref{eq:gain-par} that $g(x,0) > 0$ holds for all $0 < x < 1$, ensuring $\bar g(0) > 0$. If $c = r/\gamma_{m-1}$, we have $d_{k}(c) \le 0$ with strict inequality holding in all cases but $k = m-1$. From \eqref{eq:gain-par} this implies $g(x,r/\gamma_{m-1}) < 0$ for all $0 \le x \le 1$, ensuring $\bar g(r/\gamma_{m-1}) < 0$.

Because $\bar g(0) > 0$ and $\bar g(r/\gamma_{m-1}) < 0$ hold and $\bar{g}(c)$ is continuous the intermediate value theorem implies that there exists $0 < \bar c < r/\gamma_{m-1}$ satisfying $\bar g(\bar c) = 0$. By monotonicity of $\bar{g}(c)$ it follows that $\bar g(c) < 0$ holds for $c > \bar c$ and $\bar g(c) > 0$ holds for $c < \bar c$. The generalized version of Theorem 1 in \citet{Souza2009}  then follows from our Result~{\ref{prp:dktwosignchanges}.1 -- except that it remains to establish the existence of $0 < \bar x < 1$ such that the interior rest points satisfy $\xl < \bar x < \xr$ for all $0 < c < \bar c$. Towards this end let $\bar x$ be a solution to the problem $\max_{0 \le x \le 1} g(x, \bar c)$. As $g(0,\bar c) < 0$ and $g(1, \bar c) < 0$ holds, we have $0 < \bar x < 1$. As $g(x,c)$ is strictly decreasing in $c$, we have $g(\bar x,c) > 0$ for all $0 < c < \bar{c}$. In conjunction with $g(0, c) < 0$ and $g(1,c) < 0$ this implies that $g(x,c)$ has at least one root in the interval $(0, \bar x)$ and at least one root in the interval $(\bar x,1)$.

\newpage

\bibliographystyle{elsarticle-harv}
\bibliography{bernstein}

\clearpage

\begin{figure}
  \centering
  \includegraphics[width=\textwidth]{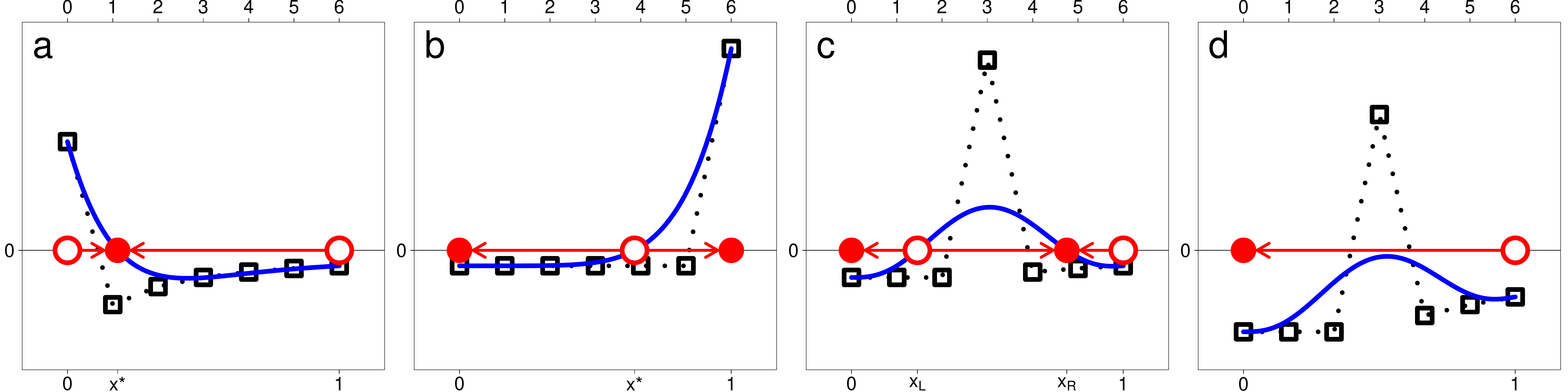}
  \caption{Gain sequence $\dkk$ (squares, dotted line; top axis), and corresponding gain function $g(x)$ (solid line; bottom axis) and phase portrait (circles, arrows) for threshold games given by~\eqref{eq:thres} and~\eqref{eq:cost_thres} with $N=7$, $r=2$, $c=1$, and
  (a) $m=1$ (see section~\ref{sec:thres-1}),
  (b) $m=N=n+1$ (see section~\ref{sec:thres-2}), or
  (c) $m=4$ (see section~\ref{sec:thres-int}).
  Panel d illustrates the same game as in panel c, but with $c=3$ instead of $c=1$.}
  \label{fig:threshold}
\end{figure}

\clearpage

\begin{figure}
  \centering
  \includegraphics[width=\textwidth]{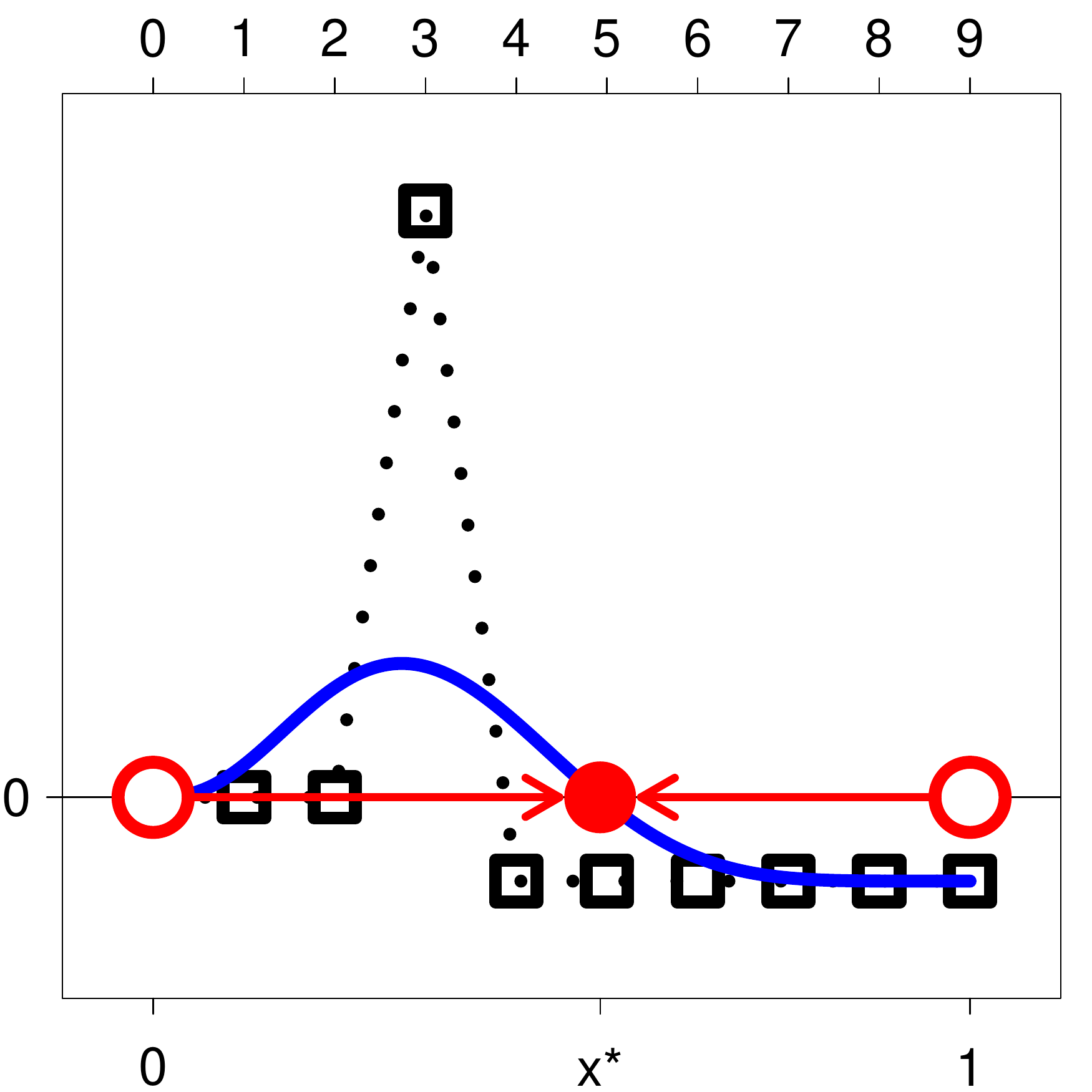}
  \caption{Gain sequence $\dkk$ (squares, dotted line; top axis), and corresponding gain function $g(x)$ (solid line; bottom axis) and phase portrait (circles, arrows) for the threshold game given by~\eqref{eq:thres-econ} with $N=10$, $r=2$, $m=4$, and $c_k=1/4$ for all $k\ge 3$.
  }
  \label{fig:hoffler}
\end{figure}

\clearpage

\begin{figure}
  \centering
  \includegraphics[width=\textwidth]{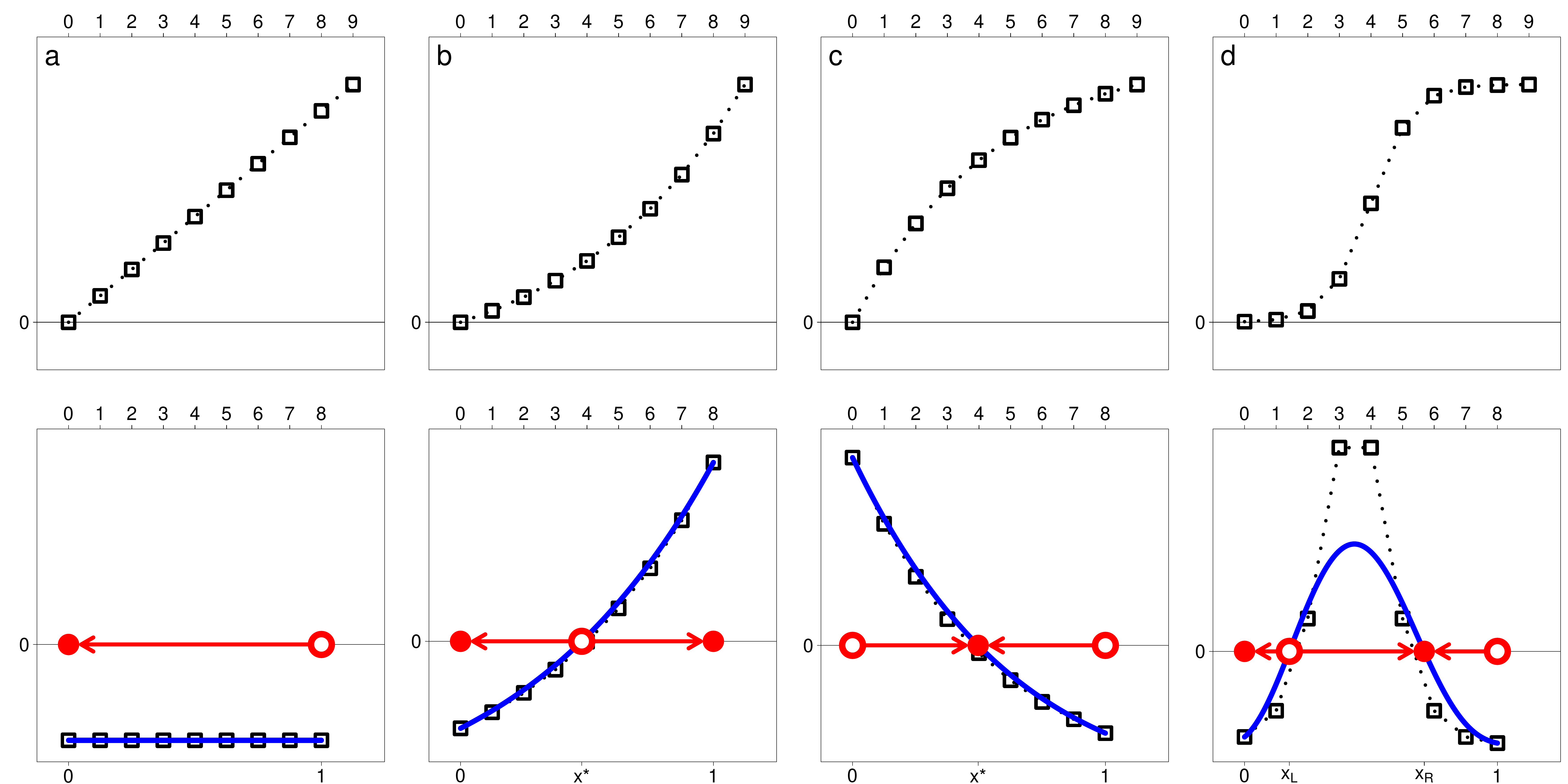}
  \caption{Examples of constant cost games with $N=n+1=9$ and $c=1/2$ for different benefit sequences.
  The first row shows the benefit sequence $r_j$; the second row shows the gain sequence $\dkk$ (squares, dotted line; top axis), and corresponding gain function $g(x)$ (solid line; bottom axis) and phase portrait (circles, arrows).
  (a) Linear benefits (see Section~\ref{sec:linpgg}) with $r=5$ and $c=1$.
  (b) Convex benefits (see Section~\ref{sec:con_benefits}) as given by~\eqref{eq:hauert-1} with $r=5$ and $w=1.2$.
  (c) Concave benefits (see Section~\ref{sec:con_benefits}) as given by~\eqref{eq:hauert-1} with $r=20$ and $w=0.8$.
  (d) Sigmoid benefits (see Section~\ref{sec:sigmoid}) as studied by~\cite{Archetti2011} with $r_j = r/[1+\exp(-s(j-m))]$, $r=20$, $m=4$, and $s=1.5$.
  }
  \label{fig:pggs}
\end{figure}

\clearpage

\begin{figure}
  \centering
  \includegraphics[width=\textwidth]{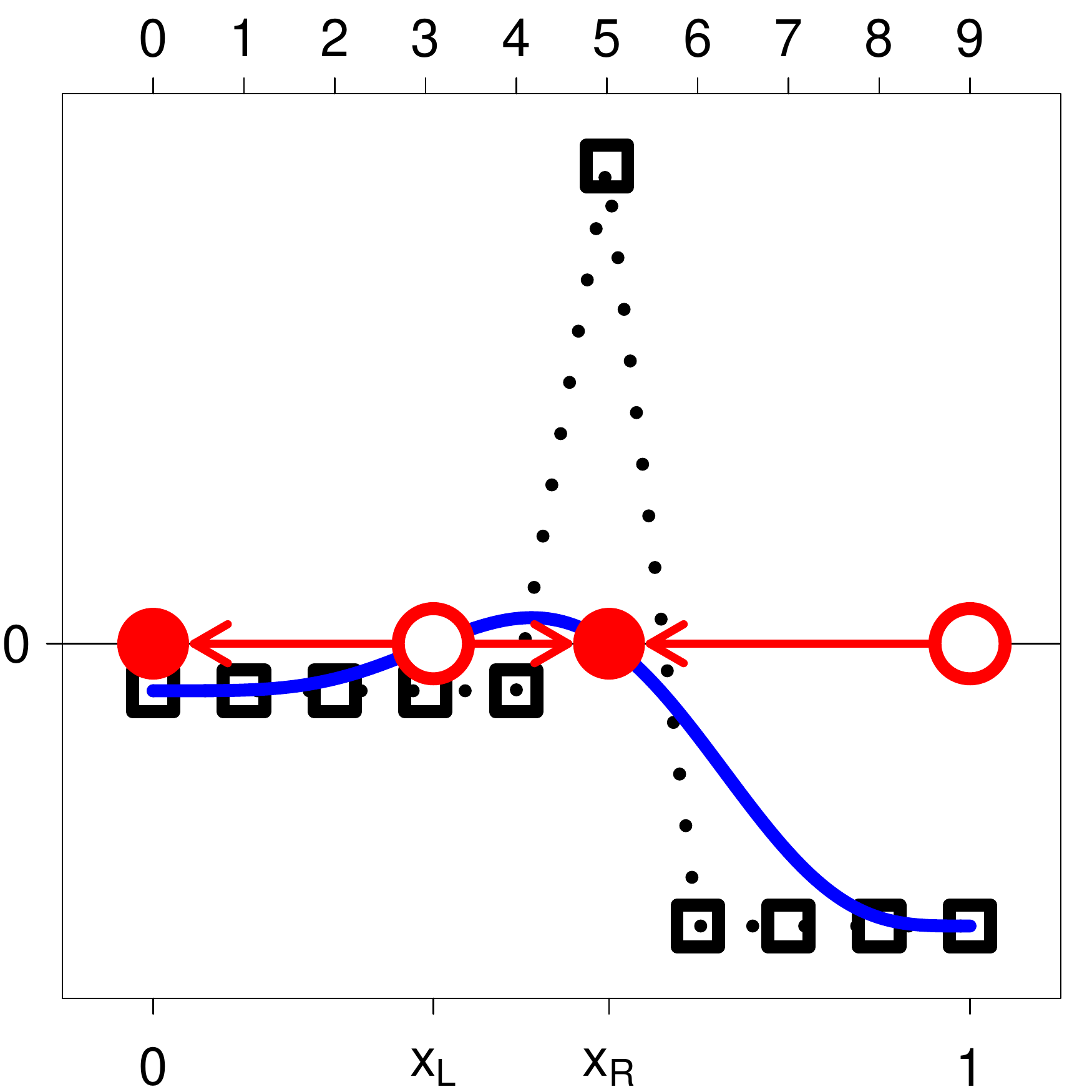}
  \caption{Gain sequence $\dkk$ (squares, dotted line; top axis), and corresponding gain function $g(x)$ (solid line; bottom axis) and phase portrait (circles, arrows)
  for the repeated $N$-person prisoner's dilemma given by~\eqref{eq:dkrnpd} with $N=10$, $r=7$, $c=2$, $T=5$, and $m=6$.
  }
  \label{fig:repeatednpd}
\end{figure}

\clearpage

\begin{figure}
  \centering
  \includegraphics[width=\textwidth]{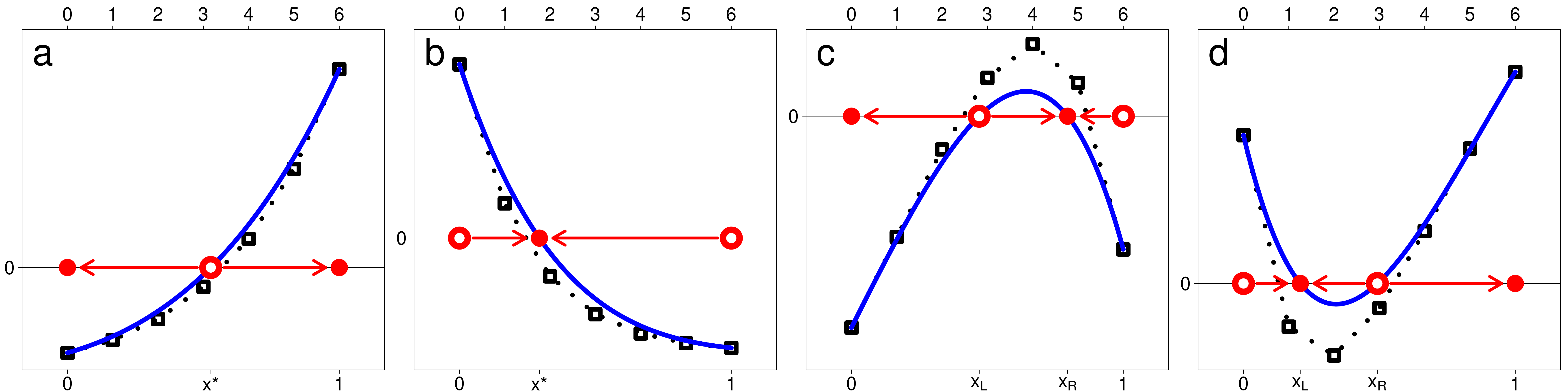}
  \caption{Gain sequence $\dkk$ (squares, dotted line; top axis), and corresponding gain function $g(x)$ (solid line; bottom axis) and phase portrait (circles, arrows) of the game considered in Section 5.2 for $N=7$
  and different values of the parameters $w$, $v$, $r$ and $c$.
  (a) $w=1.3$, $v=1.2$, $r=1$, $c=3$.
  (b) $w=0.6$, $v=0.57$, $r=2$, $c=1$.
  (c) $w=1.3$, $v=1.4$, $r=2$, $c=3.4$.
  (d) $w=0.75$, $v=0.6$, $r=1.55$, $c=1.25$.
  }
  \label{fig:hauert}
\end{figure}

\end{document}